\documentclass[showpacs,prl,letterpaper,twocolumn,aps,epsf,superscriptaddress]{revtex4}
\usepackage{dcolumn}
\usepackage{amsmath}
\usepackage{graphicx}

 \newcommand{\BF}[1]{\mbox{\boldmath $#1$}}
 
\def\sigmab{{\BF \sigma}}

\begin{document}
\title{First-order phase transition in easy-plane quantum antiferromagnets}

\author{S. Kragset} \affiliation{ Department of Physics, Norwegian
  University of Science and Technology, N-7491 Trondheim, Norway}

\author{E. Sm{\o}rgrav} \affiliation{ Department of Physics, Norwegian
  University of Science and Technology, N-7491 Trondheim, Norway}

\author{J. Hove} \affiliation{ Department of Physics, Norwegian
  University of Science and Technology, N-7491 Trondheim, Norway}

\author{F. S. Nogueira} \affiliation{Institut f{\"u}r Theoretische
  Physik, Freie Universit{\"a}t Berlin, Arnimallee 14, D-14195 Berlin,
  Germany}

\author{A. Sudb\o} \affiliation{ Department of Physics, Norwegian
  University of Science and Technology, N-7491 Trondheim, Norway}
\affiliation{Centre for Advanced Study at the Norwegian Academy
  of Science and Letters, \\
  Drammensveien 78, N-0271 Oslo, Norway}

\begin{abstract}
Quantum phase transitions in Mott insulators do not fit 
easily into the Landau-Ginzburg-Wilson paradigm. A 
recently proposed alternative to it is the so called 
deconfined quantum criticality scenario, providing 
a new paradigm for quantum phase transitions.   
In this context it has recently been proposed
that a second-order 
phase transition would occur in a two-dimensional spin 
1/2 quantum antiferromagnet in the deep 
easy-plane limit. A check of this 
conjecture is important for understanding the phase 
structure of Mott insulators. 
To this end we have performed large-scale Monte Carlo simulations on 
  an effective gauge theory for this system, including a Berry phase term that 
  projects out the $S=1/2$ sector.  
The result is a {\it first-order} phase transition, thus contradicting 
the conjecture. 
\end{abstract}

\maketitle 

The Landau--Ginzburg--Wilson (LGW) theory for phase transitions has
been an immensely successful paradigm of physics for the last $50$
years.  It is one of the cornerstones of statistical and condensed
matter physics, providing deep insight into phase transitions
\cite{Wilson-Kogut}. The standard example is the well-known
paramagnetic-ferromagnetic phase transition. Recently, examples of
phase transitions that do not fit into the LGW paradigm have been
discussed \cite{Senthil_Science_2004,Senthil_PRB_2004,Sachdev_2004}.
A prominent example is the {\it continuous} quantum phase transitions from a
N\'eel state with conventional antiferromagnetic order into a
paramagnetic valence-bond solid (VBS) state \cite{Read-Sachdev}.  In
the N\'eel state, an $SU(2)$ symmetry is broken, while in the
VBS phase translation invariance of the 
lattice is broken. The LGW paradigm does not describe this phase transition 
correctly, since it predicts a first-order phase transition in this case. 
In view of the failure of the LGW paradigm in this and other cases, a 
new scenario has recently been proposed  \cite{Senthil_Science_2004,Senthil_PRB_2004}, 
introducing the concept of {\it deconfined quantum criticality} (DQC). 
This concept applies to systems where the order parameter can be viewed as being composed 
by elementary building blocks. For instance, in the case of the N\'eel-VBS transition the 
spinons are the building blocks of the spin field. Similarly to quarks in hadrons, the 
spinons are confined in both the N\'eel and VBS phases. The DQC scenario asserts  
that the spinons are deconfined {\it only} at the critical point. This claim is based on a subtle destructive quantum interference mechanism between instantons and the Berry phase \cite{Senthil_Science_2004}.     

An attempt to provide proof of evidence for these ideas has 
recently been put forth \cite{Senthil_Science_2004}. It 
involves a deformation of the two-dimensional Heisenberg model 
into an easy-plane quantum antiferromagnet. The effective theory 
of a spin 1/2 
quantum antiferromagnet is a $O(3)$ nonlinear $\sigma$ model with a staggered 
Berry phase factor \cite{Sachdev_2004}. Such a 
nonlinear $\sigma$ model describes the fluctuations of the orientation 
${\bf n}_j$ of the order parameter. The easy-plane deformation adds a term 
proportional to $n_{zj}^2$ to the action, which explicitly breaks the 
$O(3)$ symmetry down to $U(1)$. This lower symmetry simplifies considerably 
the analysis, especially when the $CP^1$ representation ${\bf n}_j = z^*_{j a}\sigmab_{ab} z_{j b}$ 
is used, with $|z_{j1}|^2+|z_{j2}|^2=1$ due to the local constraint ${\bf n}_j^2=1$. 
The $CP^1$ representation naturally introduces a local abelian 
gauge symmetry, since ${\bf n}_j$ is invariant under the {\it local} gauge transformation 
$z_{ja}\to e^{i\theta_{aj}}z_{ja}$. A {\it deep} easy-plane deformation forces 
$n_{zj}^2\approx 0$, thus inducing the additional local constraint 
$|z_{j1}|^2\approx|z_{j2}|^2$. This allows us to write $z_{ja}=e^{i\theta_{ja}}/\sqrt{2}$.       
The requirement of local $U(1)$ gauge invariance and the deep easy-plane limit naturally leads 
to an effective 
lattice gauge theory for a quantum antiferromagnet proposed in Ref.
\cite{Sachdev_Jalabert}, which for $S=1/2$ has the lattice Lagrangian
\begin{equation}
  {\cal L}_j=-\beta\sum_{a=1}^2 
  \cos(\Delta_\mu \theta_{ja} - A_{j \mu }) -\kappa \cos(\epsilon_{\mu\nu\lambda}
\Delta_\nu A_{j\lambda})+ i \eta_j A_{j \tau}, 
\label{H-Sachdev-Jalabert}
\end{equation}
where ${\bf A}_j$ is a compact gauge field which here is doing more than just being 
an auxiliary field, like in the case of the CP$^1$ model. It determines
also the Berry phase for the above model. The index $\tau$ corresponds to imaginary time 
and the staggering factor is given by $\eta_j = (- 1)^j$. Note that the
present gauge field is a function of the spacetime coordinates, in
contrast to the usual Berry gauge potential appearing in spin
models \cite{Sachdev_2004}, which is a functional of the spin field.

Compactness of the gauge field gives rise to instanton configurations
\cite{Polyakov} which are known to spoil the  phase
transition in a corresponding model with only one phase 
field and no Berry phase \cite{Fradkin-Shenker}. In the absence of 
Berry phase and with two phase fields present, on the 
other hand, a phase transition in the 
$3DXY$ universality class occurs \cite{Babaev,Smiseth} regardless of whether the gauge
field is compact or not. Since only one gauge field is present and there are two 
phase fields available, the Higgs mechanism is able to supress only one out of two 
massless modes. The remaining massless mode is charge neutral and drives the 
$3DXY$ transition \cite{Smiseth}.  

Recently, it has been argued that the Berry phase, which is crucial to
describe the phase inside the paramagnetic phase
\cite{Senthil_Science_2004,Senthil_PRB_2004}, suppresses the
instantons at the critical point. 
Here we will investigate this point by monitoring the phase transitions 
in the model (\ref{H-Sachdev-Jalabert}) in the 
presence and absence of a Berry phase.  In the former case 
a phase transition is expected in
the charged sector, contrasting with the 
transition driven only by the neutral sector in absence of 
the Berry phase. The result will be shown to be a first-order phase transition.

The DQC scenario implies that the critical
point is governed by an easy-plane system Lagrangian featuring a
non-compact gauge field, i.e.,
\begin{equation}
{\cal L}_i =  -\beta\sum_{a=1}^2 
\cos(\Delta_\mu \theta_{ia} - A_{i \mu})
+\frac{\kappa}{2}({\bf \Delta} \times {\bf A}_i)^2.
 \label{H_noncompact}
\end{equation}
This model with unequal bare phase stiffnesses has been studied in
great detail \cite{Smiseth}. It features two distinct second-order
phase transitions, one belonging to the $3DXY$ universality class and another one corresponding to the so-called inverted $3DXY$ transition \cite{Dasgupta}. In the limit where the bare phase stiffnesses are equal, clear signals of non-$3DXY$ behavior are 
seen \cite{Smiseth,Kuklov_2005,Motrunich}. 
In Ref. \onlinecite{Kuklov_2005}, strong indications of a first-order phase transition in a loop-gas
representation of the non-compact model Eq. (\ref{H_noncompact}), were found. We will consider both Eqs. (\ref{H-Sachdev-Jalabert}) and (\ref{H_noncompact}) in detailed Monte Carlo (MC) simulations.

For performing MC simulations on the model with a Berry
phase term, it is convenient to introduce a dual representation of
the model Eq. (\ref{H-Sachdev-Jalabert}).  In such a representation the
action is real, with a Lagrangian given by \cite{Sachdev_2004}
\begin{equation}
  {\cal L}_i =  
    \frac{1}{2\beta}\sum_{a=1}^2({\bf \Delta} \times {\bf h}_i^{(a)})^2 
  +\frac{1}{2\kappa}
    ({\bf h}_i^{(1)} + {\bf h}_i^{(2)} + {\bf f}_i
    + {\bf \Delta} s_i)^2.
  \label{CL_Model}
\end{equation}
Here, ${\bf h}^{(a)}$ are integer-valued dual gauge fields, and
$\varepsilon_{\mu \lambda \nu} \Delta^{\nu} f_i^{\lambda} =
\delta_{\mu \tau} \eta_i$. Note that we would obtain Eq.
(\ref{CL_Model}) both for Eqs. (\ref{H-Sachdev-Jalabert}) and
(\ref{H_noncompact}), with $f_i = 0$ for Eq. (\ref{H_noncompact}).
For Eq. (\ref{H-Sachdev-Jalabert}) with compact 
$A_{i \mu}$, $s_i$ is integer-valued. For 
Eq. (\ref{H_noncompact}) with a non-compact $A_{i \mu}$,
$s_i$ is real-valued. Therefore in the former case $s_i$ can 
be gauged away since the ${\bf h}^{(a)}$-fields are 
integer-valued. We have chosen a gauge where $s_i = 0$.
The MC computations were performed using Eqs. (\ref{H_noncompact}) and
(\ref{CL_Model}).  For both Eqs. (\ref{H_noncompact}) and
(\ref{CL_Model}), we have used $\kappa=\beta$. We have used the
standard Metropolis algorithm with periodic boundary conditions on a
cubic lattice of size $L \times L \times L$.  For Eq. (\ref{CL_Model})
we have used $L=4,8,12,16,20,24,32,36,48,60,64,72,80,96,120$, while
for Eq. (\ref{H_noncompact}) we have used $L=48,64,80,96,112,120$. A
large number of sweeps is required in order to get adequate statistics
in the histograms (see below) for the largest system sizes. Firstly,
we have computed the second moment of the action $M_2 \equiv \langle
(S-\langle S \rangle )^2 \rangle$ for the model with and without a
Berry phase term, where $ S = \sum_i {\cal L}_i$. Secondly, we have focused on a number of quantities
that provide information on the character of the phase transition
associated with the specific heat anomaly. The first of these
quantities is the third moment of the action, $M_3 \equiv
\langle(S-\langle S \rangle)^3 \rangle$. At a second-order phase
transition this quantity should scale as follows. The peak-to-peak
height scales as $L^{(1+\alpha)/\nu}$, whereas the width between the
peaks scales as $L^{-1/\nu}$ \cite{Smiseth2003}. At a first-order
phase transition, these quantities scale as $L^6$ and $L^{-3}$,
respectively \cite{firstorderexp}. We also study the probability
distribution $P(S,L)$ of the action $S$ for various system sizes. At a
first-order phase transition, $P(S,L)$ will exhibit a double-peak
structure associated with the two coexisting phases.
 
\begin{figure}[htb]
\centerline{\scalebox{0.5}{\rotatebox{0.0}{\includegraphics{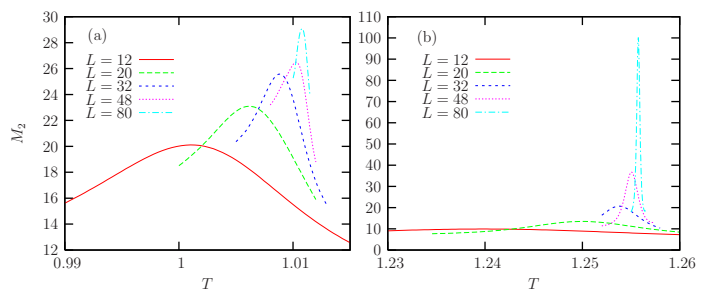}}}} 
\caption{(Color online) Specific heat $M_2$ of Eq. (\ref{CL_Model})
  for various system sizes. Panel (a): Without Berry phase term. The
  peak develops into a singularity of the $3DXY$ type. Panel (b): With
  Berry phase term. The peak develops into a $\delta$-function
  singularity with a peak scaling as $L^3$, consistent with a
  first-order transition. Note the symmetry and asymmetry of the
  peaks in the right and left panels, respectively. This is to be
  expected, since the peaks in the right panel originate with the
  superposition of a $3DXY$ peak and an inverted $3DXY$ peak. }
    \label{fig:M2}
\end{figure}

The specific heat $M_2$ is shown in Fig. \ref{fig:M2}.  Panel (a)
shows the anomaly for the model Eq. (\ref{CL_Model}) with no
Berry-phase term, i.e., ${\bf \Delta} \times {\bf f} = (0,0,\eta) =
0$.  The anomaly has the characteristic asymmetric shape of
the $3DXY$ model. In this case, there are no Berry phases to suppress
the instantons of the compact gauge-field ${\bf A}$ at the critical
point. Hence, the charged sector does not feature critical
fluctuations that can interfere with those of the neutral sector. When
the Berry phase field ${\bf f}$ is included, the specific heat is
notably more symmetric and the anomaly develops into a
$\delta$-function peak, consistent with a first-order phase
transition. This is shown in panel (b).

\begin{figure}[htb]
\centerline{\scalebox{.65}{\rotatebox{0.0}{\includegraphics{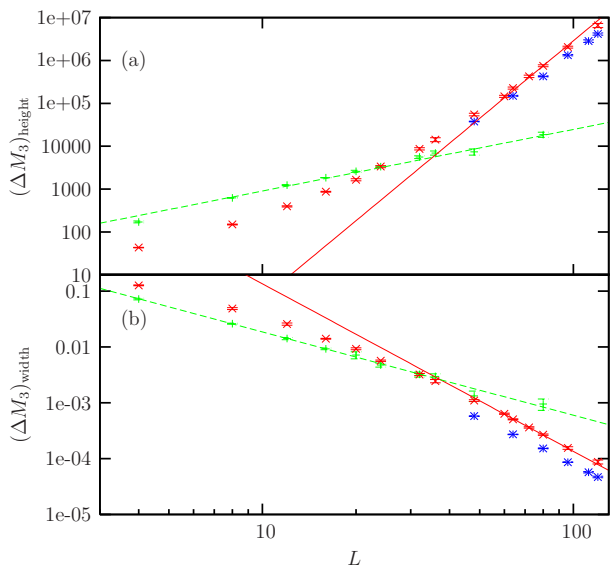}}}} 
\caption{(Color online) Scaling of the height [panel (a)] and width
  [panel (b)] of $M_3$ of the action in Eqs.  (\ref{H_noncompact}) and
  (\ref{CL_Model}).  The lines in panel (a) represent $L^{1.43}$ and
  $L^6$. The former is the $3DXY$ result. The lines in panel (b)
  represent $L^{-1.49}$ and $L^{-3}$. The former is the $3DXY$ result.
  For large system sizes, the height and width scale in manner
  consistent with a first-order phase transition. Also shown are
  results for Eq. (\ref{CL_Model}) with no Berry phase term ${\bf
    f}=0$ (green symbols). These results follow the $3DXY$ scaling
  lines. The red symbols are the results for Eq. (\ref{CL_Model})
  while the blue symbols are results for Eq. (\ref{H_noncompact}).}
    \label{fig:M3scale}
\end{figure}

To investigate more precisely the character of the phase 
transition when a Berry
phase term is present, we have performed finite-size scaling (FSS) of
the third moment of the action, $M_3$ \cite{Smiseth2003}.  The results
are shown in Fig. \ref{fig:M3scale}, panel (a).
It is seen that for small and intermediate system sizes, the height
increases with $L$ in a manner which might appear consistent with that
of a second-order phase transition. However, the quality of the
scaling is not satisfactory, since a clear curvature in the scaling
plots is seen (red data points). As system sizes increase we see a
gradual increase in the apparent value of $(1+\alpha)/\nu$, until for
large system sizes, we clearly have $M_3 \sim L^6$, consistent with a
first-order phase transition \cite{firstorderexp}.

Panel (b) of Fig. \ref{fig:M3scale} shows the scaling of the width of
$M_3$. Again, the line with the smallest negative slope is the line
one would obtain for the $3DXY$ model, while the line with the most
negative slope is $\sim L^{-3}$, characteristic of a first-order phase
transition. Again we obtain apparent scaling, with a crossover regime
at intermediate length scales into a regime where the width scales as
it would in a first-order phase transition \cite{firstorderexp}. The
results of Fig. \ref{fig:M3scale} provide further support to the
notion that the phase transition in the model with a compact gauge
field and a Berry phase term is a first-order phase transition.

\begin{figure}[htb]
  \centerline{\scalebox{.65}{\rotatebox{0.0}{\includegraphics{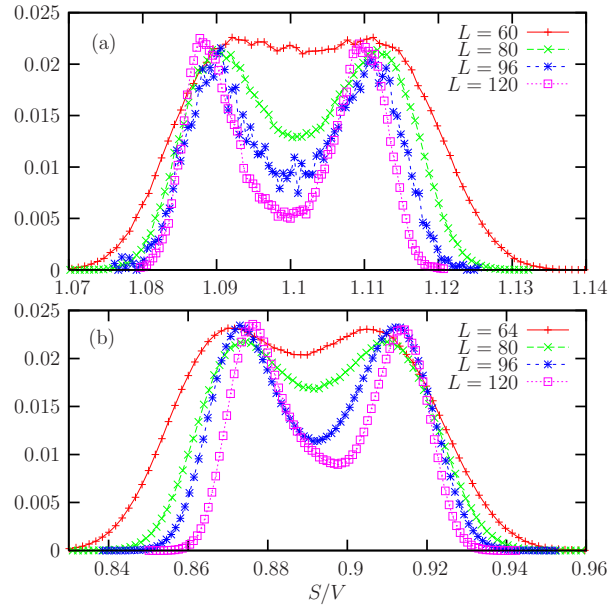}}}}
  \caption{(Color online) 
     Histograms for the probability distribution
    $P(S,L)$ as a function of $S/L^3$ for various system sizes 
    $L$. (a): results for 
    Eq. (\ref{H-Sachdev-Jalabert}) in the representation 
    Eq. (\ref{CL_Model}). (b):  results 
    Eq. (\ref{H_noncompact}). A double peak structure
    develops with the latent heat per unit volume 
    approaching a finite constant as $L$ is increased.  
    This is a hallmark of a first-order
    transition.  For the largest systems, 
    up to $120 \cdot 10^6$
    sweeps over the lattice were performed. A total of approximately
    $500 000$ CPU hours were used to obtain these results.}
    \label{fig:histogram}
\end{figure}

To investigate this further, we have computed the probability
distribution $P(S,L)$ for various system sizes.  The results are shown
in Fig. \ref{fig:histogram}. Panel (a) shows results for Eq.
(\ref{H-Sachdev-Jalabert}) in the representation Eq. (\ref{CL_Model}).
Panel (b) shows results for Eq. (\ref{H_noncompact}). The
Ferrenberg--Swendsen algorithm has been used to reweight the histograms
\cite{Ferrenberg-Swendsen}. For $L \leq 48$, we essentially have not
been able to resolve a double peak structure at all, showing that the
phase transitions in the models Eqs. (\ref{H-Sachdev-Jalabert}) and
(\ref{H_noncompact}) are weakly first-order. We have located the
transition temperature from the peak structures in the specific heat
and $M_3$, and performed long simulations at this temperature for each
$L$. For the largest systems, $L=96, 120$, up to $120 \cdot 10^6$
sweeps over the lattice were done. A clear double-peak structure
in $P(S,L)$ is seen to develop for system sizes $L > 60$. The fact
that such large system sizes are required to bring out the double-peak
structure, implies that this phase transition is weakly first-order.

\begin{figure}[htb]
\centerline{\scalebox{.65}{\rotatebox{0.0}{\includegraphics{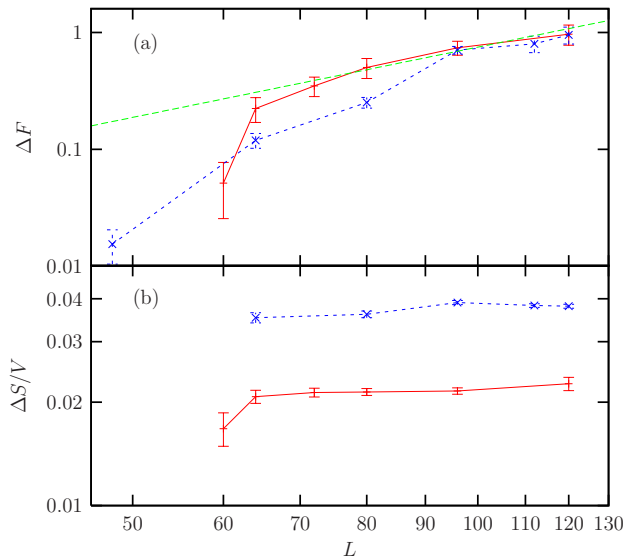}}}} 
\caption{(Color online) Panel (a) shows the scaling of the height
  $\Delta F$ of the peak between the two minima in $-\ln P(S,L)$ both
  for Eqs. (\ref{CL_Model}) (red curve) and (\ref{H_noncompact}) (blue
  curve). Dotted line is the line $\sim L^2$.  The height 
  scales as $\Delta F \sim L^{d-1}$. This is
  a hallmark of a first-order transition. Panel (b) shows
  latent heat  per unit volume $\Delta S/V$ as a function of $L$.
  The upper
  (blue) curve is from Eq. (\ref{H_noncompact}), the lower (red)
  curve is from Eq. (\ref{CL_Model}). 
  $\Delta S/V$ approaches a nonzero value as $L$ is increased.}
    \label{fig:DF}
\end{figure}

We also perform FSS of the height of the peak between the two
degenerate minima in the free energy $-\ln [ P(S,L) ] $.  This height
should scale as $L^2$ in a first-order phase transition, since it
represents the energy of an area which separates two coexisting phases
\cite{Lee-Kosterlitz}.  The results are shown in panel (a) of Fig.
\ref{fig:DF}.  For large enough systems, the height clearly approaches
the dotted line $\sim L^2$, as in a first-order transition.  This is
corroborated by extracting the latent heat per unit volume in the
transition, shown in the lower panel of Fig. \ref{fig:DF}. It
approaches a nonzero constant as $L$ is increased, as it should in a
first-order phase transition.

Further insight into the nature of the first-order phase transition
can be obtained by means of the renormalization group (RG). In the
field theory Lagrangian the interaction of an easy-plane system reads
${\cal L}_{\rm int}=u_0(|z_1|^2+|z_2|^2)^2/2+v_0|z_1|^2|z_2|^2
=u_0(|z_1|^4+|z_2|^4)/2+w_0|z_1|^2|z_2|^2$, where $w_0=u_0+v_0$.
Consider a generalized situation where the complex fields have each 
$N/2$ components. The renormalized
dimensionless couplings in $d=4-\varepsilon$ dimensions are 
$g=u\mu^{-\varepsilon}$, $h=w\mu^{-\varepsilon}$, and $f$, where $f$
is the dimensionless gauge coupling and $\mu$ is an arbitrary mass
scale. The $\beta$ functions at one-loop order are
\cite{Flavio_RG_Paper} $\beta_g=-\varepsilon
g-6gf+(N+8)g^2/2+2Nh^2+6f^2$, $\beta_h=-\varepsilon
h-6hf+3(N+2)gh+6f^2$, and $\beta_f=-\varepsilon f+Nf^2/3$. 
Nontrivial fixed points
with $f=3\varepsilon/N$ and $h<0$ are found for $N\geq 300$, while in the 
deep easy-plane limit $h=0$ {\it no fixed points with 
$f=3\varepsilon/N$ are found for all values of $N$}.   
In a Ginzburg-Landau (GL) theory of superconductors, the
existence of a critical value of $N$ above which nontrivial 
fixed points are found actually reflects the strong-coupling behavior at much 
lower values of $N$. It turns out that the
phase transition for one complex order parameter is second-order in the type II regime
\cite{Dasgupta}, while a first-order transition occurs in the
type I regime \cite{Kleinert-tric,Sudbo-tric}. Inspired by the GL case,   
we interpret the complete absence of a critical value of $N$ for $h=0$  
as a clear signature of a first-order phase transition in the 
deep easy-plane regime. This is a further confirmation of our MC results;  
see also Ref. \cite{Kuklov_2005}, where a first-order transition 
in a closely related model has also been found. 

In summary, our large scale MC simulations of the deep easy-plane 
quantum antiferromagnet confirm the instanton-Berry phase 
suppression mechanism proposed in Ref. \cite{Senthil_Science_2004}. Therefore, 
the spinons are indeed deconfined at 
the phase transition. However, the phase transition in this case is first-order, which contradicts the DQC picture for this model, where a second-order phase transition has been predicted.    
    
This work was supported by the Research Council of Norway, Grant Nos. 158518/431, 158547/431 (NANOMAT), and 167498/V30 (STORFORSK), 
and the Norwegian High-Performance Computing Consortium
(NOTUR).

\end{document}